A Need for Dedicated Outreach Expertise and Online Programming: Astro2020 Science White Paper

**Authors:** Amanda E Bauer (AURA/LSST, abauer@lsst.org), Britt Lundgren (UNC Asheville), William O'Mullane (AURA/LSST), Lauren Corlies (AURA/LSST), Megan E. Schwamb (Gemini Observatory), Brian Nord (FermiLab), Dara J Norman (NOAO)

**Endorsers:** Laura Trouille, Zooniverse co-PI, VP of Citizen Science, The Adler Planetarium, Northwestern University, Cameron Hummels, Caltech, Joshua Pepper, Lehigh University, Ranpal Gill, AURA/LSST, Andrés Plazas, Princeton/LSST, Douglas A. Caldwell, SETI Institute, Adrian Price-Whelan, Flatiron Institute, Jennifer Sobeck, UW/SDSS, Christine O'Donnell, University of Arizona, Robert Blum, AURA/LSST, Phil Marshall, SLAC/LSST, Mark Newhouse, AURA/NOAO, Kim Coble, San Francisco State University

*Type of Activity: Infrastructure Activity State of the Profession Consideration Other Outreach, Education, Communications*

## Summary (350 character limit):

Maximizing the public impact of astronomy projects in the next decade requires NSF-funded centers to support the development of online, mobile-friendly outreach and education activities. EPO teams with astronomy, education, and web development expertise should be in place to build accessible programs at scale and support astronomers doing outreach.

## 1. Executive Summary and Recommendations

The purpose of large astronomy projects is to educate and inspire the public, as they are the primary stakeholders in all government-funded activities. Therefore, the astronomy community and NSF-funded centers need to capitalize on positive trends in digital literacy, the increasing use of mobile devices, and a discovery space driven by social media, through the progressive development of effective online resources in astronomy education and public outreach (EPO).

**EPO teams who develop leading-edge activities will have started the design process as early as possible, including during construction of new facilities. To build data-driven activities for non-specialists using modern web technologies that are accessible, interesting, engaging, and effective at scale, EPO teams will need to draw on a number of areas of expertise: astronomical research methods, educational theory and practice, web development and design, software engineering, and multi-modal communication.**

The following recommendations clarify what needs to be done to maximize the impact of ongoing outreach efforts, bolster the opportunities that exist within modern web tools, and elevate society's engagement with the process of scientific discovery in the decade of petabyte-scale astronomy.

RECOMMENDATION 1: Create accessible online activities for the public. To maximize the impact of astronomy in society in the rapidly approaching petabyte and exabyte eras, we recommend that projects, facilities, and data centers develop accessible web interfaces and tools that enable the public to access, explore, and analyze authentic astronomical data at unprecedented scale, within the framework of an appropriate narrative.

RECOMMENDATION 2: Bring dedicated/domain experts onto astronomy education and outreach teams. To create the accessible online interfaces that are appropriate for non-specialists and maximize public impact, we recommend supporting dedicated education and outreach teams that bring together astronomers, technical experts, and education specialists to strengthen accessibility, relevance, and usability of activities.

RECOMMENDATION 3: Fund dedicated or centralized astronomy education and outreach groups. We recommend that funding agencies supporting the development and operation of large astronomical observing and data facilities to fund professional education and outreach groups who can provide strategy, oversight, and effective practices to maximize the impact of outreach efforts, incorporate EPO efforts as part of a project's mission, and amplify overall agency investment in *Broader Impacts*.

## 2. Create accessible online activities for the public

To prepare for the EPO response to the large data we will collect in the 2020s, it is critical to examine existing examples of online astronomy EPO programs. Consistent themes emerge from this exercise: it will be critical to produce well-defined learning outcomes for activities, provide curated access to authentic data, and develop simple, intuitive web interfaces. There is a necessity for all activities to be presented through mobile-friendly interfaces, and additionally, that materials for non-specialists are presented in a way that require no software installation or downloading of data (especially important for educators introducing new activities to their classrooms).

RECOMMENDATION 1 (repeated): Create accessible online activities for the public. To maximize the impact of astronomy in society in the rapidly approaching petabyte and exabyte eras, we recommend that projects, facilities, and data centers develop accessible web interfaces and tools that enable the public to access, explore, and analyze authentic astronomical data at unprecedented scale, within the framework of a narrative appropriate for non-specialist audiences.

Traditional means of public engagement (e.g. classroom visits, online videos, public lectures and panels, etc.) have demonstrated their importance and value, and have carved a niche in the landscape of public engagement. However, we have entered a new era of technology and social interaction, which necessitates new modalities for innovative pedagogical techniques, communication, and even scientific exploration. Many good arguments have been made for enabling non-professionals and students to access and engage with authentic data and professional tools. However, in practice, the increasing complexity of interfaces to large datasets can become a barrier to access and use.

User interfaces need to be attractive and intuitive for non-specialists and usable from mobile devices and platforms commonly used in schools (such as chromebooks and tablets). Interfaces created for professionals do not necessarily work for non-specialists, because they tend to have the following characteristics: 1) offer too many options; 2) do not offer a clear path toward a learning outcome; 3) are too slow, unresponsive, or burdensome for the internet connections. Effort should be spent on user interfaces for public audiences, and ideally, on creating introductory activities as preparation for more complicated tasks. It's also the case that interfaces created for the public can work well for professionals as "quick look" tools.

Surveying users to assess their needs and interests helps the content design process and continues to improve the quality of an experience for users when a program in running. User testing is a regular practice for many companies that deliver a product to the public and is a process that should be adapted within astronomy EPO programs to ensure activities remain relevant and useable.

The remaining sections identify areas of expertise EPO teams can employ to achieve these outcomes, and describe examples of existing online astronomy outreach programs.

## 3. Bring expertise to astronomy education and outreach teams

RECOMMENDATION 2 (repeated): Bring dedicated/domain experts onto astronomy education and outreach teams. To create the accessible online interfaces that are appropriate for non-specialists and maximize public impact, we recommend supporting dedicated education and outreach teams that bring together astronomers, technical experts, and education specialists to strengthen accessibility, relevance, and usability of activities.

Maintaining support for astronomy research relies on our ability to effectively communicate our science to a variety of audiences and cultivate public excitement and engagement. Historically, strategic programming for astronomy EPO within projects has been primarily undertaken by individual astronomers who are passionate about EPO yet are rarely compensated for their time or rewarded by their effort in EPO and science communication. While these individual efforts are beneficial and impactful, astronomers are not expected to know effective practices around mobile-friendly development, how to develop intuitive user interfaces for the public, marketing through social media, or how to connect astronomy activities to formal education curriculum standards. A team of EPO experts can advise and assist with these areas, which are essential to build successful activities that are discoverable, impactful, adoptable, and accessible at scale.

We recommend astronomy organizations support creating EPO teams with expertise in relevant areas. This section describes options for areas of expertise and roles that can be brought on to achieve specific goals.

An **Astronomical Analysis Specialist** could bring astronomical expertise to the team, act as a liaison to professional astronomers who consult with the team, and could bring data visualization and data analysis skills to the team.

Educators in the US are currently required to submit paperwork to demonstrate that they are teaching specific topics related to curriculum standards.  An EPO **education specialist or instructional designer** brings knowledge of relevant curriculum standards and rubrics (for example, the Next Generation Science Standards[1]) and is able to connect astronomy activities to topics educators must cover. This is the most relevant for K-12 formal education in a traditional setting or homeschooling. An educational specialist can build professional development programs to increase confidence for bringing such activities into their classrooms if there is not an expert available to join in person.

An education specialist can also tap into educator networks to advertise existing programs and perform professional development. An example is the National Science Teachers Association (NSTA) annual meeting and AAS. An education specialist working with an astronomer can create curate datasets to achieve specific learning outcomes without overwhelming non-specialists \citep{Rebull18}.

An **Information Officer or Communications Manager** can act as a primary contact for a facility and also set overall communication strategies and implementation plans. Topics covered in such strategies could include audiences, content priorities, communication channels, messaging, procedures, and more.

An **Outreach Specialist** could serve a range of purposes depending on the needs of the group.  This could be a science writer, someone who responds to and directs questions received from audience groups, or contributes to social media presence. or an astronomer trained in science communication. If this person has astronomy training, he/she could work with astronomical datasets to curate options for the public.

A **web developer** considers the user interface and experience when visiting a site. Mobile-friendly accessibility is a requirement for non-specialists since most users of an online interface will discover the materials via social media and will access them from a mobile device, not a desktop platform. In addition, the most common machines used by schools are chromebooks and potentially weak internet connections, which require lightweight design. Development needs to satisfy these requirements and produce interactivity are best implemented by experts in the field.

Social Media is becoming increasingly important as a source of news and information in society. Dedicating a full-time equivalent (or more) to the role of **social**

---

[1] https://www.nextgenscience.org/

**media engagement specialist** increases awareness of EPO activities and engages various audiences to participate with activities that exist.

Overall branding, the look and feel of online activities, and developing interesting graphics and images to support press releases or other activities are the role of a **Graphic Designer**. There can also be an element of experience in marketing strategies and brand recognition within this type of role.

An **evaluation specialist** informs methods for understanding the impact of programs on specific audience groups. The most benefit occurs when the method for evaluating the success of a program is built into the development of the program itself. Metrics could include and are not limited to web analytics, short or long surveys, interviews, login requests, focus groups, web submission forms, and social media interactions.

A **Software Architect** designs, deploys, and maintains production services for an online program. It is important to not overburden internet systems that can be common in classroom settings or non-urban areas.

A **Project Manager** oversees the detailed budget, schedule, contracts, documentation, and reporting. This role is important for programs being built during the construction of an astronomical facility.

## 4. Fund dedicated astronomy education and outreach groups

RECOMMENDATION 3 (repeated): Fund dedicated or centralized astronomy education and outreach groups. We recommend that funding agencies supporting the development and operation of large astronomical observing and data facilities fund professional education and outreach groups who can provide strategy, oversight, and effective practices to maximize the impact of outreach efforts, incorporate EPO efforts as part of a project's mission, and amplify overall agency investment in *Broader Impacts*.

Having a dedicated individual or team to develop the EPO program for a specific facility can improve efficiency, impact, and cost effectiveness. Strategic planning provides an opportunity to emphasize the identity of a particular large facility; to identify non-specialist audiences who could benefit the most from dedicated engagement; put into place best practices in outreach and communication programs; and complement the overall landscape of astronomy EPO efforts. It is important that the EPO professionals are employed directly at professional telescope facilities in order to emphasize the uniqueness of the program, build and maintain relationships with those doing the technical and scientific work, and help handle the astronomy-specific data products that currently require a reasonable level of understanding to interpret and use.

A dedicated EPO team also serves as a resource for enabling astronomers working with large datasets and data facilities to do more impactful and wide-reaching outreach. Groups that are specifically charged to do EPO can improve the impact of the existing NSF *Broader Impacts*[2] investment by supporting astronomers to tap into existing programs, thus improving the overall quality of outreach that is being done. This improves discoverability of the EPO work astronomers are doing increases the likelihood of achieving Broader Impact goals at both the individual and NSF levels.

An EPO team could provide any of the following benefits:

- Conducting science communication and media training sessions for astronomers doing these activities.
- Providing introductions to various social media platforms that can be used for unique outreach experiences.
- Marketing and promoting activities through established social media and common online training resources (e.g. Code Academy).
- Creating or tapping into a centralized repository for people looking for resources
- Creating opportunities for collaboration between astronomers and existing outreach infrastructure that will promote success and provide wide-reaching impact. Examples include *Journey Through the Universe* in Hawai'i or *AstroDay* in Chile, both led by Gemini EPO.
- Performing user needs assessments and user testing to improve the quality of existing activities and to develop new programs that meet the needs of specific audiences.
- Evaluating and reporting on the impact of EPO activities. Evaluation methods can and should be built into program design.
- Providing guidance for astronomers when developing science drivers and use cases for educational materials and public interfaces related to their research expertise.
- Broadening participation to non-traditional audience groups.

The timing for building expert EPO teams should occur during the construction of a new facility and be included as part of the project's mission. Starting early affords time to implement appropriate strategy and infrastructure. Educational materials, supplemental professional development materials, striking visualizations and images, and communications strategies should be ready at the start of a project to maximize the public impact of the facility.

## Examples of existing online activities

Several examples of existing and planned infrastructures illuminate avenues for online public engagement: below, we discuss Sloan Digital Sky Survey's (SDSS) SkyServer, Zooniverse's Citizen Science, and the EPO program of LSST.

---

[2] https://www.nsf.gov/od/oia/special/broaderimpacts/

For over 15 years, the SDSS has made its vast database freely accessible to the world via the web-based SDSS data browser, SkyServer[3]. The query and analysis tools available through the SkyServer are designed to meet the needs of astronomers and non-professionals alike.  However, the large number of available features and the technical jargon that accompany them can overwhelm non-experts, as well as professional astronomers who are external to the SDSS collaboration.

In order to better support non-specialist audiences, the SDSS Education and Public Outreach team developed activities with simplified query tools and smaller, curated datasets to facilitate activities for pre-college educators and students (e.g. SDSS *Voyages*[4]). For non-specialist audiences, these activities lower the barrier to accessing the same authentic data, while providing an introduction to concepts related to both astronomy and data structures. For students and educators who may be interested in using the data for more advanced explorations, SDSS Voyages provides a helpful stepping stone to the full functionality of the SkyServer.

Citizen science represents an example of successful use of the modern age of web connectivity by directly engaging the public in scientific research. Online citizen science enables scientists to work with the general public to perform data-sorting and analysis tasks that are difficult or impossible to automate, or that would be insurmountable for a single person or for small groups of individuals to undertake [1]. Most participants from the public claim that the main reason they participate is the contribution they are making to fundamental science research [2]. Through online citizen science portals such as the Zooniverse[5] [3], millions of volunteers have participated directly in this collaborative research experience, contributing to over 70 astronomy-based research papers. Another reason for the continued success of the Zooniverse platform in particular, is that it looks good and feels modern, even after a decade of activity. While professional astronomers are the PI's of citizen science projects, Zooniverse employs 13 developers, one designer, and two postdocs to lead the infrastructure development of the platform between the Adler Planetarium and Oxford, UK locations.

Members of the Zooniverse team have furthered the project's educational impact by developing a college-level data science curriculum around their crowd-sourced data. The NSF-funded Improving Undergraduate STEM Education (IUSE) Project: "Engaging Introductory Astronomy Students in Authentic Research through Citizen Science" (PI: L. Trouille) is a particularly successful example of scoping big-data astronomy for a college non-major audience. This innovative curriculum equips students with freely available web-based tools to explore the intrinsic and environmental properties of a large, curated sample of  SDSS galaxies with morphological classifications from Galaxy Zoo.  The curriculum empowers students to explore authentic data without requiring full-scale datasets or jargon-rich professional tools for visualization and analysis.

---

[3] http://skyserver.sdss.org

[4] http://voyages.sdss.org

[5] http://www.zooniverse.org

LSST's EPO program[6] is unique among ground-based telescope projects: not only is it being constructed in tandem with the physical observatory itself, but the outreach program is funded at 2% of the project cost. EPO products will go live when the LSST Survey begins in 2022 and were included from the beginning as part of the construction Project deliverables. Some major findings from early user needs assessments include the necessity for mobile-friendly interfaces, a clear path toward learning objectives through narratives, and educators needing no new software to download in order to introduce classroom activities. This has shaped the overall strategy for LSST EPO development and the skill sets needed on the EPO Team, which is a small, interdisciplinary team of astronomers, writers, designers, educators, and developers.

The LSST EPO team is developing online, data-driven classroom investigation activities for students in advanced middle school through college. The topics cover commonly-taught principles in astronomy and physics, and each investigation is designed for use with Next Generation Science Standards (NGSS) in the United States and the Curriculum Nacional in Chile. All investigations come with support and assessment materials for instructors and no special software is needed to access the investigations, which will be available in English and Spanish. Finally, LSST EPO will provide support to researchers who create Citizen Science projects using LSST data, including a dedicated project-building tool on the Zooniverse platform. The infrastructure to host these activities is being built during construction and will take several years. Another critical task during construction is building prototypes and performing user testing, which has continually proven to improve the user experience and usability of interfaces.

## Conclusion

In this paper, we discussed recommendations and effective practices that can be employed to maximize the impact of large astronomy facilities and data centers in the next decade. We prefaced the need for creating accessible online astronomy activities for the public and identified a range of skills needed to create such activities. Finally, we established the benefits of resourcing dedicated EPO groups from the earliest stages of astronomy facility planning and even including EPO as part of the mission of projects.

These recommendations are driven by two factors: trends seen elsewhere on the web that successfully respond to the modern era of technology and social interactions on the web, and case studies within astronomy that demonstrate appropriate avenues for increasing engagement and accessibility through online activities.

[1] "Ideas for Citizen Science in Astronomy," Marshall, P. J., Lintott, C. J., & Fletcher, L. N., 2015, ARA&A, 53, 247,
https://www.annualreviews.org/doi/10.1146/annurev-astro-081913-035959

---

[6] https://www.lsst.org/about/epo